\documentclass[aps,prb,preprint,groupedaddress]{revtex4}
\usepackage{graphicx}
\usepackage{epstopdf}
\begin{document}

\title{Phonon propagation dynamics in band-engineered\\one-dimensional phononic crystal waveguides}

\author{Daiki Hatanaka}

\email{hatanaka.daiki@lab.ntt.co.jp}

\author{Amaury Dodel}

\author{Imran Mahboob}

\author{Koji Onomitsu}

\author{Hiroshi Yamaguchi}

\affiliation{NTT Basic Research Laboratories, NTT Corporation, Atsugi-shi, Kanagawa 243-0198, Japan}

\begin{abstract}
The phonon propagation dynamics in a phononic crystal waveguide, realized via a suspended one-dimensional membrane array with periodic air holes, is investigated as function of its geometry. The bandstructure of the phononic crystal can be engineered by modifying the characteristics of the phonon standing waves in the waveguide by varying the waveguide width and the pitch of the air holes. This enables the phonon transmission bands, the bandgaps, the velocity and the nonlinear dispersion in the phononic crystal to be controlled. Indeed the engineered bandstructure can also be tuned to sustain multiple phonon modes in a given branch which whilst being spectrally degenerate can be temporally resolved via their differing group velocities. This systematic study reveals the key geometric parameters that enable the phonon transport in phononic crystal waveguides to be fully controlled. 
\end{abstract}

\maketitle

\section{INTRODUCTION}

\hspace*{0em}Phononic crystals (PnCs) have emerged as novel platform in which the phonon propagation dynamics can be controlled \cite{maldovan_thermocrystal, martinez_pnc, liu_pnc, khelif_pnc, benchabane_pnc, mohammadi_pnc1, mohammadi_pnc2, mohammadi_pnc4, otsuka_pnc, liang_pnc, boechler_pnc, hatanaka_wg, mohammadi_pnc3, maldovan_prl}. PnCs are usually composed from a periodically modulated elastic media with the period determined by the wavelength of the corresponding phonon waves. The long range periodic modulation in these structures results in phonon standing waves to be established that give rise to phonon bandgaps. This concept was first put into practice in the macroscopic regime within a two-dimensional (2D) periodic arrangement of stainless steel cylinders which could block sound waves at frequencies corresponding to the bandgap \cite{martinez_pnc}. The ultimate evolution of this concept has led to the development of 2D PnCs which consist of a 2D periodic arrangement of air holes typically in a silicon crystal \cite{mohammadi_pnc1, mohammadi_pnc2, mohammadi_pnc4, otsuka_pnc, mohammadi_pnc3}. The 2D PnC can not only completely block phonons at the bandgap frequencies but the introduction of point and line defects, by modulating the air holes, even enables phonons to be guided and trapped within these structures. Indeed this platform can control different flavors of phonons i.e. ultra-/hyper-sound \cite{mohammadi_pnc3, mohammadi_pnc2, mohammadi_pnc4} and even heat \cite{maldovan_thermocrystal, maldovan_prl} and their high confinement efficiencies have enabled concepts such as phonon multiplexing to be established for signal processing applications \cite{mohammadi_pnc2}. The key to accessing such advanced functions is underpinned by the phonon dispersion relation in these structures which can be engineered by careful design of the 2D periodic air hole matrix.\\
\hspace*{1em}More recently, a variation of 2D PnCs has been developed in PnC waveguides (WGs) where phonons are confined and guided via a one-dimensional (1D) array of suspended electromechanical membranes \cite{hatanaka_wg}. In contrast to 2D PnC WGs, where phonons are confined within the line defect defined WG via the phonon bandgap of the surrounding 2D PnC, the phonon confinement in the 1D PnC WG is achieved by the acoustic impedance mismatch between the suspended membrane and the local bulk environment. However, the periodic air hole array used to suspend the membranes imparts a periodic elastic potential that is experienced by the phonon confined within the WG and it results in the emergence of phononic bands, bandgaps, slow phonons and even nonlinear dispersion which can broaden the phonon packets in the WG. However in spite of their richness, a detailed physical understanding of these effects and how they can be harnessed, akin to 2D PnCs, remains elusive. Indeed such knowledge holds the tantalizing promise of greatly enhancing the functionality of the PnC WG architecture. For instance, photonic crystals represent the apex of such engineered structures in which photon dispersion relations are designed by tuning their periodic refractive index structure \cite{yablo_ptc, krauss_ptc, joan_ptc, akahane_q, notomi_slow, baba_ptc, engelen_ptc, millan_fwm, monat_fwm, colman_soliton, redondo_soliton}. This in turn enables a whole range of optical phenomenon to be controllably accessed including four wave mixing \cite{millan_fwm, monat_fwm}, supercontinuum generation \cite{dudley_sc1, dudley_sc2} and even soliton formation \cite{colman_soliton, redondo_soliton, russell_soliton} for a multitude of applications.\\
\hspace*{1em}To that end, the purpose of this study is to elucidate the key physical parameters of the PnC WG architecture that enables the phonon dispersion relations to be designed for a specific application. These systematic experimental investigations, underpinned by finite element method (FEM) analysis, reveal that by adjusting both the waveguide and air hole geometry enables the phononic transmission bands, phonon bandgaps, phonon velocities and the nonlinear dispersion to be fully controlled in the PnC WG. Furthermore, these studies not only show the availability of energy degenerate phonon bands, which can be resolved in temporal measurements, but they also indicate the conditions needed to selectively activate them thus adding a completely new and hitherto unconsidered functionality to the PnC architecture.

\section{EXPERIMENT}

\hspace*{0em}The 1D PnCs reported in this study are shown in Fig. 1(a) and were synthesized from an epitaxially grown single crystal GaAs/AlGaAs heterostructure as detailed in Fig. 1(b). From this substrate, 3 PnC WGs with different hole pitches ($\it a$) of 8, 10 and 12 $\mu$m were fabricated on the same chip as shown in Fig. 1(c). In total 3 chips were prepared each sustaining 3 WGs of fixed widths (${\it w}$) of 22, 29 and 34 $\mu$m, which yielded 9 different PnC WGs that were investigated.\\
\hspace*{1em}Specifically, the PnC WGs were fabricated by first patterning a mesa structure on the heterostructure via photolithography that was subsequently wet etched using a H$_{3}$PO$_{4}$(6\%):H$_{2}$O$_{2}$(6\%):H$_{2}$O(88\%) solution. The mesa structure was exposed to a height of approximately 500 nm in order to eliminate electrical cross-talk between the neighboring PnC WGs through the Si-doped GaAs layer. Next, electrodes were patterned at the edges of the WGs via photolithography and were deposited with gold using the lift-off technique. The 1D air hole array was then defined along the [110] crystal orientation where the number of holes were chosen to fix the length of all PnC WGs to approximately 1.0 mm. The holes with a radius of 2.5 $\mu$m were also patterned with photolithography and were then wet etched all the way down to the Al$_{0.65}$Ga$_{0.35}$As layer with H$_{3}$PO$_{4}$:H$_{2}$O$_{2}$:H$_{2}$O solution. Finally, each chip was immersed in HF(5$\%$):H$_{2}$O(95\%) solution to isotropically etch the Al$_{0.65}$Ga$_{0.35}$As sacrificial layer which yielded the suspended membrane array that composed the PnC WG \cite{hatanaka_wg, hatanaka_bis, hatanaka_memarray}. By varying the HF immersion time between 25, 33 and 40 minutes, 3 chips sustaining 3 PnC WGs with $\it w$ = 22, 29 and 34 $\mu$m were prepared. In order to ensure high quality mechanical performance, all the chips were also cleaned post fabrication via an ultraviolet and O$_{3}$ dry stripper to remove any residual contamination on the surface of the WGs.\\
\hspace*{1em}The experiments were then performed at room temperature and in high vacuum ($\sim$ 2 $\times$ 10$^{-4}$ Pa) where the phonons were piezoelectrically generated in the WGs with the application of alternating voltage between the gold electrode and the Si-doped GaAs layer from a signal generator (NF WF1974) as shown in Fig. 1(a). The resultant vibrations, confined to the WG, were optically detected with He-Ne laser Doppler interferometer (Neoark MLD-230V-200). The electrical output from the interferometer's photodetector was measured either in a vector signal analyzer (HP 89410A) or in an oscilloscope (Agilent DSO6014A) in which case the output from the interferometer was first preamplified (NF SA-220F5).

\section{RESULTS AND DISCUSSIONS}
\subsection{Spectral response}

\hspace*{0em}First, in order to evaluate the bandstructure as function of the PnC geometry, the transmission of all 9 PnC WGs were spectrally measured via the protocol depicted in Fig. 1(a). The spectral responses of 3 WGs with (${\it w}$, ${\it a}$) = (22, 10), (29, 12) and (34, 8) in microns corresponding to devices 1, 2 and 3 henceforth are shown in Figs. 2(a)-2(c) respectively. In addition, the corresponding dispersion relations extracted from FEM simulations (COMSOL Multiphysics) are also shown in these figures. In device 1, phonon propagation in the WG emerges above a cut-off frequency of 3.8 MHz which corresponds to the 1st dispersion branch from mode A detailed in Fig. 2(d). However, as the WG width increases in devices 2 and 3, the cut-off frequency reduces to 2.2 and 1.6 MHz as shown in Figs. 2(b) and 2(c). Physically this trend arises from the larger membranes composing the WGs in these devices sustaining lower frequency resonances \cite{hong_memb}. Indeed the inverse dependence of the cut-off frequency to the WG width is observed irrespective of the hole pitch and is reproduced by the FEM simulations as shown in Fig. 3(a).\\
\hspace*{1em}The transmission measurements in device 1 (Fig. 2(a)) also reveal a phonon bandgap at 6 MHz that is created when the half wavelength of the phonon vibrations in the WG corresponds to the hole pitch length.  This results in Bragg reflection that leads to standing waves which suppress phonon propagation at the corresponding frequencies. Meanwhile, the FEM analysis reveal that this bandgap corresponds to the 1st and 3rd dispersion branches that sustain mode shapes with an antinode (A in Fig. 2(d)) or a node (B in Fig. 2(d)) between the holes in the WG with the same wavelength. In other words, the WG can support 2 vibrations of the same wavelength but different energies corresponding to the different mode profiles. On the other hand, as the hole pitch is increased to 12 $\mu$m in device 2 shown in Fig. 2(b), the bandgap between the 1st and 3rd branches closes due to the greater overlap between their corresponding mode shapes as a consequence of their longer wavelengths. Consequently, this observation naturally suggests that a shorter hole pitch will yield shorter wavelengths for the phonon modes in the 1st and 3rd branches giving rise to less overlap which should result in a more distinct bandgap emerging. However, experimentally decreasing the hole pitch to 8 $\mu$m in device 3 shown in Fig. 2(c) reveals a much less pronounced phonon bandgap. This unexpected observation originates from the 3rd dispersion branch in this device sustaining mode shapes labeled D (see Fig. 2(d)) that overlaps with the 1st bandgap at the Brillouin zone boundary and thus mask it as revealed by the FEM simulations in Fig. 2(c).\\
\hspace*{1em}The dispersion relations extracted from the FEM simulations, shown in Figs. 2(a)-2(c) indicate that the PnC WGs can sustain a range of phonon branches whose corresponding mode shapes are detailed in Fig. 2(d). However, the phonon modes corresponding to the 2nd and 4th branches (gray lines in Figs. 2(a)-2(c)) arising from mode shapes labeled C and E in Fig. 2(d) make a negligible contribution in the present study due to the piezoelectric transducer's inability to activate them as a consequence of the electrodes being located at their nodal position. In contrast, the 5th branch sustains mode shapes that exhibit tension mediated intermodal interactions with the modes in the 3rd branch that results in their anti-crossing which opens up a new bandgap around 7 to 9 MHz in devices 2 and 3 as shown in Figs. 2(b) and 2(c) \cite{imran_eit, okamoto_coupling}. Although the spectral transmission measurements are unable to clearly resolve these higher order bandgaps as they are located far from the Brillouin zone boundary, surprisingly the temporal measurements detailed below can do so. 

\subsection{Phononic bandgap}

\hspace*{0em}The spectral transmission measurements in all 9 PnC WGs enable both the cut-off frequency (described above) and the phonon bandgap to be systematically evaluated as function of the WG's geometry and are summarized in Figs. 3(a) and 3(b) respectively.\\
\hspace*{1em}This analysis unsurprisingly reveals that the spectral positions of the 1st bandgap is strongly dependent on the WG width and the hole pitch length as shown in Fig. 3(b). Specifically, reducing the hole separation gives rise to shorter wavelength standing waves in the WG thus yielding a bandgap at higher frequencies, whereas the WG width determines the spectral position of all the phonon branches as referenced to the cut-off frequency detailed in Fig. 3(a). Indeed the FEM simulations confirm the former trend as shown by the solid line in Fig. 3(b) for a PnC WG with ${\it w}$ = 29 $\mu$m whilst ${\it a}$ is continuously varied from 7 to 13 $\mu$m. The spectral width of the 1st bandgap also shows an inverse correlation with the hole separation. However this trend is not captured by the widest WG with ${\it a}$ = 8 $\mu$m as a consequence of the emergence of the 3rd branch which overlaps the 1st bandgap as described above.\\
\hspace*{1em}Somewhat surprisingly, the spectral position of the 2nd bandgap also reveals a stronger than expected correlation on the hole separation even though this bandgap arises from the interaction between the phonon modes in the 3rd and 5th branches. However as the hole pitch is reduced for a given PnC WG width, all the phonon branches migrate to higher frequencies which consequently results in the anti-crossing between the 3rd and 5th branches also occurring at higher frequencies. Again this trend is confirmed by the FEM analysis as shown by the dotted lines in Fig. 3(b).\\
\hspace*{1em}Consequently these systematic analyses reveal that the spectral positions of the bandgaps and somewhat generally their spectral widths can be tailored by simply controlling the hole pitch in the PnC WGs.

\subsection{Temporal response}

\hspace{0em}In order to gain a deeper understanding of the phonon propagation dynamics in the 1D PnC WGs, temporal measurements are also performed. In Fig. 4, the spatially mapped phonon propagation in device 2 in response to a rectangular shaped input pulse of 2 $\mu$s duration is shown. This measurement is executed with a pulse frequency of 5.6 MHz at which phonon modes sustained by the 3rd dispersion branch are activated as detailed in Fig. 2(b). The mechanical vibrations corresponding to these phonons are excited from one edge of the PnC WG and they subsequently propagate through the WG with a constant speed of 134 m/s and a loss of 0.4 dB/mm. Once these phonons reach the opposite edge of the WG, they are reflected, which results in a Fabry-Perot resonance that can be identified by the equally spaced peaks in the spectral measurements as highlighted in Fig. 2(b).\\
\hspace*{1em}Next to evaluate the temporal characteristics of the phonon propagation in the WG with respect to the underlying bandstructure, similar time of flight measurements are performed as function of frequency in all 3 devices. Phonons in the PnC WGs are excited by rectangular input pulses from the WG's right edge with durations of 2 $\mu$s and 8 $\mu$s which are then only detected at the left edge, in contrast to the spatial mapping measurements detailed above, and are shown in Figs. 5(a)-5(c). As expected, phonons in all 3 devices could only be observed above the cut-off frequency irrespective of the pulse duration. Once the phonons reached the opposite edge of a WG they are reflected resulting in an enhanced signal (highlighted in Fig. 4). Indeed multiple reflections occur as identified by the enhanced intensity fringes in all 3 devices. Although the phonons propagate with constant velocity in the WGs but when the pulse frequencies approach the edges of the bandgap, phonon slowing occurs namely the phonon group velocity (${\it v}$$_{\rm g}$) is reduced. This can be clearly seen by the change in curvature of the reflection fringes as highlighted by the black arrows in Figs. 5(a)-5(c). The physical origin of this effect arises from a change in the curvature of a dispersion branch with respect to wavevector at frequencies around the bandgap \cite{notomi_slow, baba_ptc, hatanaka_wg}. Indeed this group velocity dispersion (GVD) can also broaden the phonon pulse packet as it travels down the PnC WG as detailed below.\\
\hspace*{1em}In contrast to devices 1 and 2 where the extremities of the 1st bandgap could be identified from the phonon slowing, such unambiguous identification of the bandgap in device 3 was not possible. In fact the temporal measurements indicate the existence of an interference like response in this spectral region which is also observed in the spectral measurements depicted in the inset to Fig. 2(c). These observations can be understood from the FEM simulations (also shown in Fig. 2(c)) which indicate that the 3rd dispersion branch overlaps with the 1st bandgap. Consequently the suppression of modes A and B (see Fig. 2(d)) in parallel with the availability of mode D (see Fig. 2(d)) and the subsequent overlap of their modal profiles gives rise to the complex temporal response observed in this device for frequencies in and around the bandgap. It should be noted that in contrast to the spectral measurements, the bandgap region in device 3 could be more easily identified in the temporal measurements due to this highly pronounced interference effect.

\subsection{Group velocity}

\hspace*{0em}The time of flight measurements detailed above also enable the group velocity to be quantitatively extracted. Specifically, the first fringe corresponding to a one way trip in the PnC WG is Gaussian fitted at a given frequency. The group velocity at this frequency can then be readily determined from the ratio of the WG's length to the time obtained from the center of the Gaussian fit. Extending this analysis to all the frequencies displayed in Figs. 5(a)-5(c) yields the spectral dependence of the group velocity in all 3 PnC WGs as shown in Figs. 6(a)-6(c).\\
\hspace*{1em}First, the extracted group velocities show no dependence on the pulse width where clearly the narrower pulse corresponds to a spectrally broader frequency packet and vice versa. Second, in the 1st and 3rd dispersion branches (away from the 1st bandgap and only for mode shapes A and B in Fig. 2(d)) spanning 3 different frequency ranges of 3.5-10 MHz, 2-7 MHz and 2-6 MHz in devices 1, 2 and 3 respectively, almost identical group velocities are obtained. Indeed the group velocities determined directly from the FEM simulated bandstructure shown in Figs. 2(a)-2(c), from the relation $v_{\rm g} = 2\pi(df/dk)$, where the ${\it f}$ and ${\it k}$ are frequency and wavenumber respectively, not only reproduce the experimentally observed group velocities but they also confirm that the phonon velocities are inextricably linked to these dispersion branches. Finally, the extracted group velocities not only exhibit the slowing effect in the proximity of the 1st bandgap, but it can also be seen around the 2nd bandgap (see Figs. 6(b) and 6(c)) even though these bandgaps cannot be resolved in the spectral measurements described in Figs. 2(b)-2(c). These results also unsurprisingly indicate that the broader 8 $\mu$s pulse, corresponding to a spectrally pure input, can probe the rapidly changing dispersion branches in the spectral regions around the bandgaps with a higher resolution than the 2 $\mu$s pulse.\\
\hspace*{1em}The suppression in the group velocity can also modify the temporal waveforms of the pulse fringes as depicted in Figs. 6(d)-6(f). This effect can be attributed to the GVD which can be spectrally identified from the second derivative of the FEM simulated dispersion branches in Figs. 2(a)-2(c), namely the GVD coefficient $k_{2} = (1/2\pi)^2(d^2k/df^2)$ as shown in the upper panels of Figs. 6(a)-6(c) \cite{agrawal, engelen_ptc}. As $k_{2}$ increases in the proximity of the cut-off and the 1st bandgap at 4.3 MHz in devices 1 and 2 respectively (see Figs. 6(d) and 6(e)), the reflection fringes are both broadened and distorted. In the proximity of the band edges where $k_{2}$ is maximized, this results in the velocity of phonons rapidly changing which experimentally manifest itself in the broadened temporal waveforms. In contrast, this frequency corresponds to the middle of the 1st dispersion branch located away from the band edges  for device 3 (see Fig. 6(c)) where $k_{2}$ is negligible and thus the corresponding group velocity is more explicitly defined. As a result the reflection fringes at this frequency also exhibit clearly defined waveforms without any broadening as shown in Fig. 6(f). Similar observations are also made in devices 1 and 2 at 7.5 MHz as detailed in Fig. 6(d) and 6(e). Consequently, designing the phonon bandstructure by simply tuning the geometry of the PnC WG enables not only the phonon group velocity at given frequency to be selectively adjusted but the corresponding waveforms can also be modified on demand via the nonlinear dispersion.\\
\hspace*{1em}Most interestingly, the temporal measurements shown in Fig. 5(c) clearly reveal the existence of 2 different reflection fringe profiles with different curvatures above 6 MHz (where this new fringe is highlighted in pink). Since this effect is most clearly observed with the 2 $\mu$s input pulse (left panel in Fig. 5(c) and lower panel in Fig. 6(f)), their temporal waveforms are selectively fitted with 2 different Gaussians. This analysis reveals the existence of 2 different group velocity profiles above 6 MHz in device 3 as shown in Fig. 6(c). As described above, at these frequencies the 3rd dispersion branch interacts with the 5th dispersion branch and the resultant anti-crossing modifies the 3rd branch so that it can sustain 2 distinct phonon modes (B and D shown in Fig. 2(d)) with different wavevectors but at the same frequency. Indeed the above analysis indicates that the additional fringe structure (see lower panel in Fig. 6(f)) yields slower group velocities which correspond to phonons in the 3rd branch whose mode profile is described by D (see Fig. 2(d)). On the other hand, the faster group velocities at these frequencies correspond to phonons whose mode shape is described by B in Fig. 2(d) from the 3rd branch as in devices 1 and 2.\\
\hspace*{1em}The combination of temporal measurements in the PnC WGs with the FEM derived bandstructure provides a powerful approach to investigating phonon dynamics in this platform. Indeed these investigations have even opened up the path to identifying different phonon modes within a particular dispersion branch. This new degree of freedom, where different vibration modes are available at the same frequency, is akin to engineering the orbital angular momentum of photons to increase the information throughput at the same frequency in optical fibers \cite{wang_mom_fiber, roland_mdm}. Consequently, these investigations bring forth the exciting prospect of geometrically engineering PnC WGs so that their bandstructure can sustain multiple vibration modes at the same frequency which in turn can be identified from temporal measurements via for example their different group velocities. If such functionality can be harnessed in the PnC WG architecture it will not only make the concept of phonon circuits more tantalizing but it will also open up new directions in phononics.

\section{CONCLUSION}

\hspace*{0em}The 1D PnC WG architecture is systematically investigated as function of its geometry via a combination of spectral and temporal measurements in conjunction with its bandstructure extracted from FEM simulations. These investigations reveal that the cut-off frequency, the spectral position and width of a given phonon band and the bandgap in the PnC WG can be controlled by simply varying the width of the WG and the pitch of the holes used to suspend the membranes. Consequently, the speed of phonons and even their temporal profile can be controlled via the GVD at given frequency by appropriately engineering the WG geometry in accordance to the bounds extracted in this report. Furthermore, the FEM simulations also indicate that phonons in only certain dispersion branches can be activated as consequence of the position of the piezoelectric transducers on the PnC WG. Indeed this analysis provides design directions in future devices where multiple piezoelectric transducers will be incorporated to selectively activate phonons in a particular dispersion branch. Most unexpectedly, the tension induced interaction between dispersion branches not only results in the creation of new bandgaps but the highly modified branches also sustained different wavevector vibration modes which coexist at the same frequency. Although the spectral measurements could not resolve these degenerate states, the temporal measurements uniquely could discriminate between the coexisting phonon modes in a given dispersion branch via their differing group velocities.\\
\hspace*{1em}Consequently, this study vividly indicates that the combination of the FEM simulated bandstructure in combination with temporal measurements are the key to ascertaining detailed knowledge of the phonon dynamics in the PnC WG architecture. Ultimately these investigations have yielded insights into the relationship of the WG's geometric parameters to its underlying bandstructure which can now be harnessed to engineer and exploit both the well-known phonon dynamics and the more exotic phononic degrees of freedom arising from the nontrivial bandstructure in PnC WGs.

\begin{acknowledgements}
We thank Y. Ishikawa for growing the heterostructure. This work was partly supported by JSPS KAKENHI Grant Number 23241046.
\end{acknowledgements}

%\bibliography{ref_band}

\newpage

\begin{figure*}[floatfix]
\begin{center}
\vspace{-5cm}\hspace{8cm}
\includegraphics[scale=0.85]{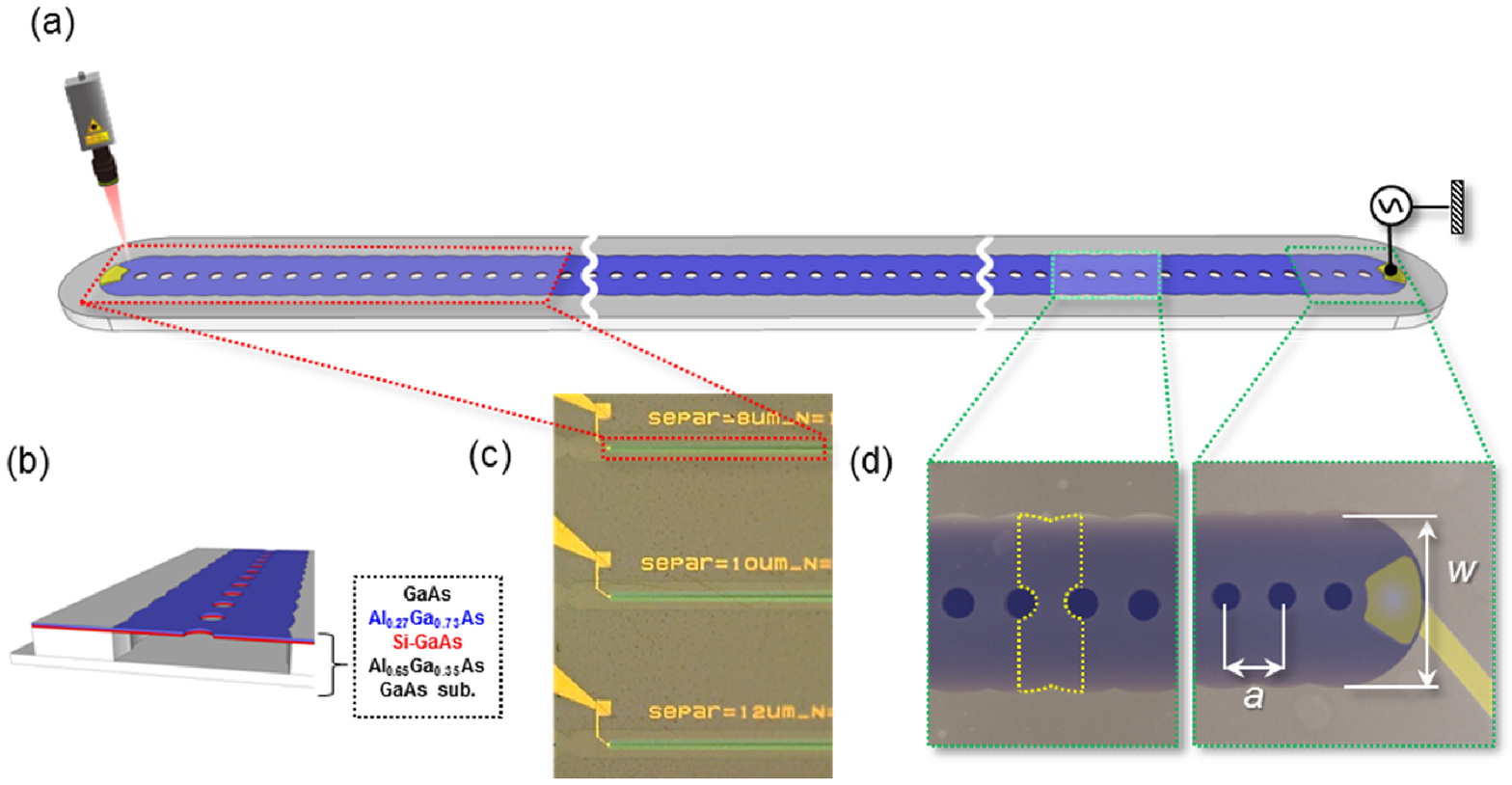}
\vspace{-11cm}
\caption{(Color online) (a) A schematic of the 1D PnC WG and the measurement set-up used to investigate phonon transmission. The piezoelectrically-active membranes composing the PnC WG (blue) are suspended through the 1D air hole array. (b) A cross-sectional schematic of the PnC WG detailing the layer profile of the GaAs (5 nm) / Al$_{0.27}$Ga$_{0.73}$As (95 nm) / Si-doped GaAs (100 nm) / Al$_{0.65}$Ga$_{0.35}$As (3 $\mu$m) heterostructure. (c) An optical photograph of one of the ${\it chips}$ containing 3 PnC WGs with different hole pitches but identical widths. (d) A false color electron micrograph of the 1D PnC WG indicating the hole pitch (${\it a}$) and the WG width (${\it w}$) namely the parameters used to define the geometry of a given WG. A unit cell in the PnC WG is also highlighted by the yellow dotted line which is used in the FEM bandstructure simulations detailed in Fig. 2(d).}
\label{fig 1}
\vspace{-0.5cm}
\end{center}
\end{figure*}

\begin{figure*}[floatfix]
\begin{center}
\vspace{-1.5cm}\hspace{-2cm}
\includegraphics[scale=0.8]{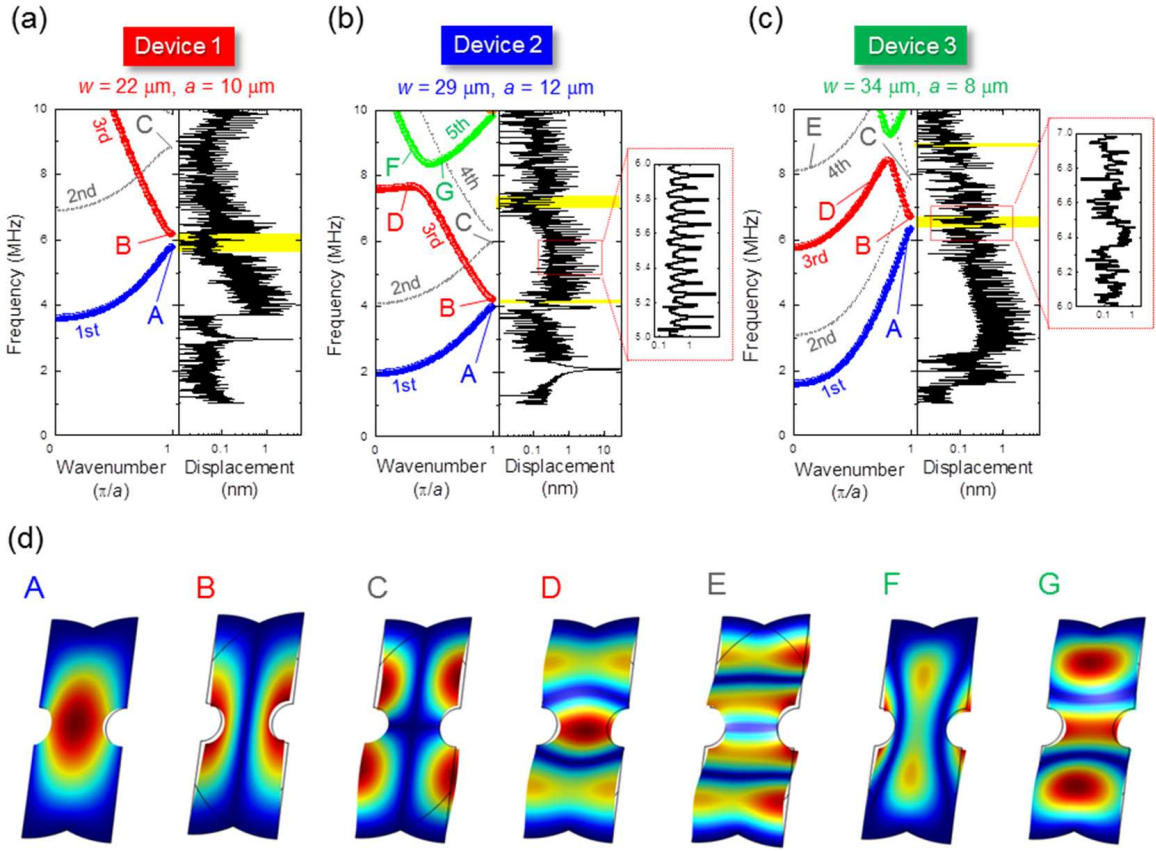}
\vspace{-9.5cm}
\caption{(Color online) (a)-(c) The FEM simulated dispersion relations of the different phonon modes (schematically detailed in (d)) sustained by the 1D PnC WGs (left panels) where the different dispersion branches are labeled. Also shown are the experimentally measured frequency responses (right panels) acquired in the transmission configuration detailed in Fig. 1(a) when excited with amplitudes of 0.5, 1.8 and 1.0 V$_{\rm rms}$ in the 3 devices respectively. The FEM simulations also incorporate the residual stress in the GaAs/AlGaAs heterostructure arising from the lattice mismatch \cite{liu_stress_gaas}. The resonance peaks around 3 MHz and 2.1 MHz in (a) and (b) are derived from vibrations localized to the WG's edge which don't contribute to the mobile phonons that propagate down to the WG. The spectral positions of the 1st and 2nd bandgaps are also highlighted in yellow. (d) The FEM derived spatial profiles of the various phonon modes sustained by the PnC WG corresponding to the different branches of the bandstructure shown above.}
\label{fig 2}
\vspace{-0.5cm}
\end{center}
\end{figure*}

\begin{figure*}[floatfix]
\begin{center}
\vspace{-3.5cm}\hspace{-2cm}
\includegraphics[scale=0.7]{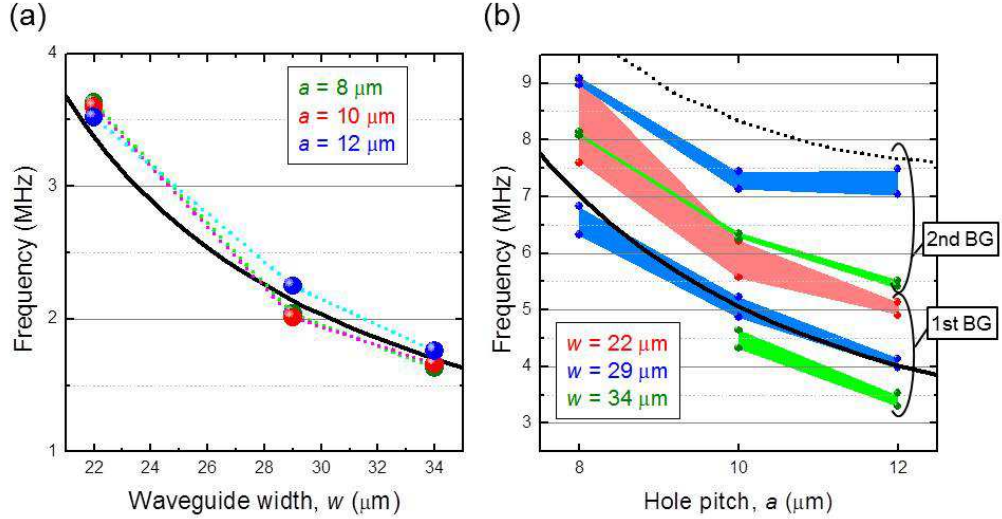}
\vspace{-5.5cm}
\caption{(Color online) (a) The spectrally measured cut-off frequency in all 9 PnC WGs exhibits an inverse correlation with the WG width for a given hole pitch (spheres with dotted lines). The cut-off frequency extracted from the FEM simulations with $a = 12 \mu$m as a function of $w$ confirms this experimental observation (solid line). (b) The spectrally measured 1st and 2nd (only with $w =$ 29 and 34 $\mu$m due to the 10 MHz measurement bandwidth in the present experimental set up) phononic bandgaps in all 9 PnC WGs. The colored spheres with shading indicate both the spectral edges and widths of the phononic bandgaps. The spectral positions of the FEM derived bandgaps (solid and dotted lines for the 1st and 2nd bandgaps respectively) with $w = 29 \mu$m as function of the hole pitch again confirm the experimental observations.}
\label{fig 3}
\vspace{-0.5cm}
\end{center}
\end{figure*}

\begin{figure*}[floatfix]
\begin{center}
\vspace{-2.8cm}\hspace{8cm}
\includegraphics[scale=0.7]{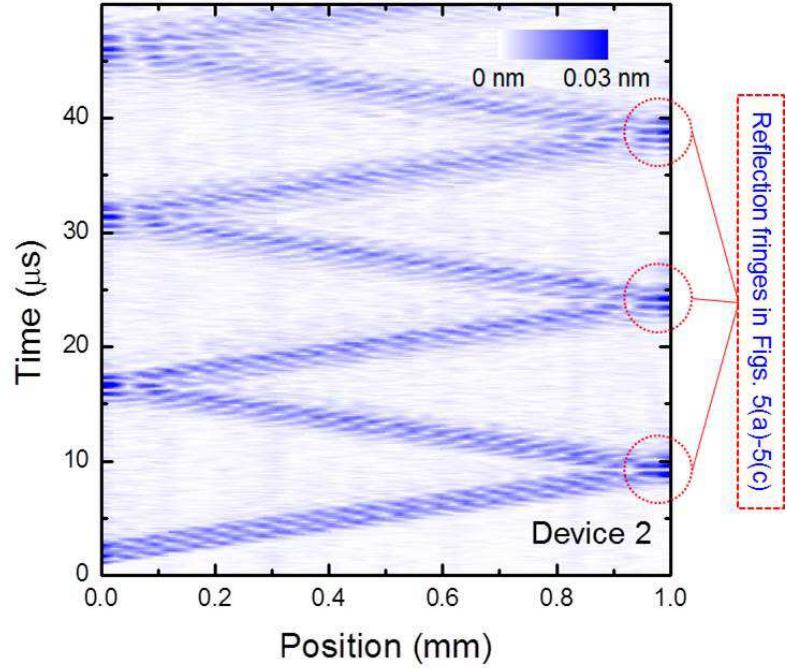}
\vspace{-5.5cm}
\caption{(Color online) The spatially mapped phonon transmission in device 2 with a pulse width of 2 $\mu$s at 5.6 MHz with an amplitude of 0.3 V$_{\rm rms}$ corresponding to phonon modes in the 3rd dispersion branch i.e. B in Fig. 2(d). The interferometer is scanned along the PnC WG which reveals that the phonons travel with constant velocity before being reflected multiple times at the WG's edges.}
\label{fig 4}
\vspace{-0.5cm}
\end{center}
\end{figure*}

\begin{figure*}[floatfix]
\begin{center}
\vspace{-1.5cm}\hspace{3cm}
\includegraphics[scale=0.7]{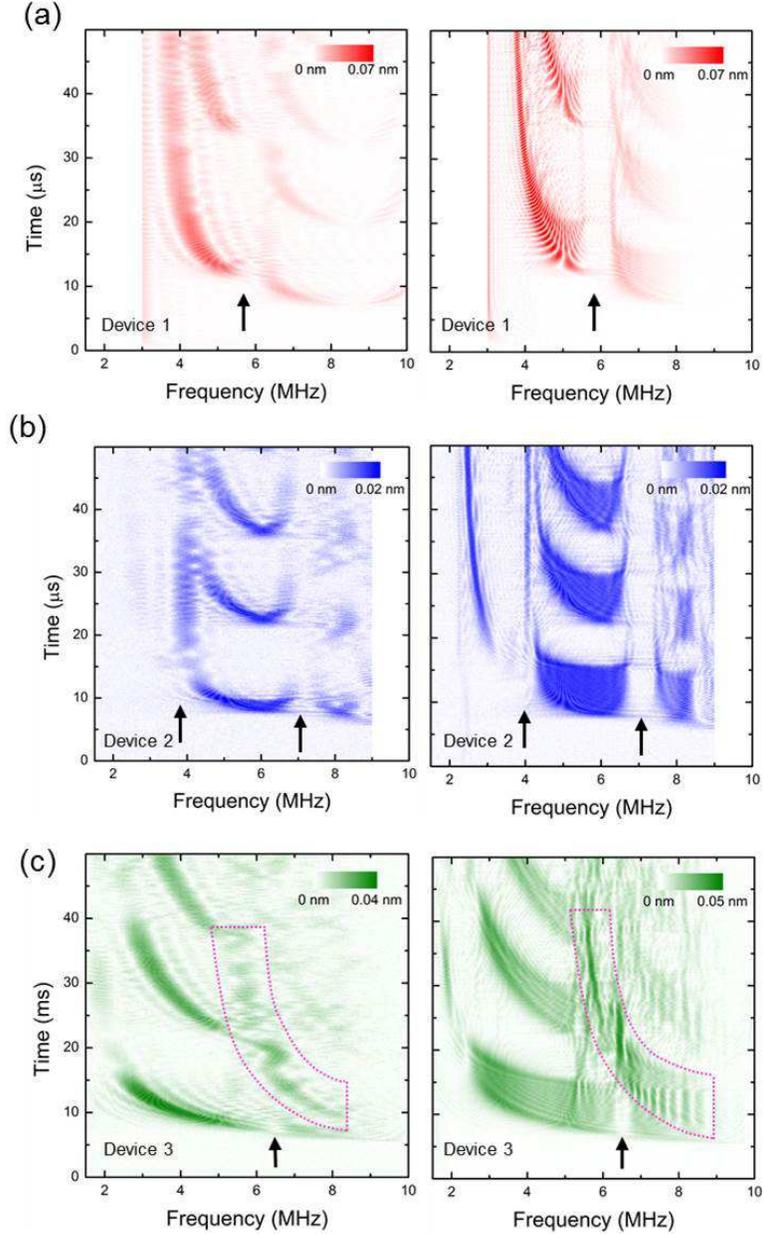}
\vspace{-1.7cm}
\caption{(Color online) (a)-(c) The temporal response of the phonon propagation in the PnC WGs when excited with a pulse of 2 $\mu$s (left panel) and 8 $\mu$s (right panel) duration with an amplitude of 0.5 V$_{\rm rms}$ in devices 1, 2 and 3 respectively. The black arrows indicate the positions of the bandgaps. The region highlighted in pink in (c) indicates the pulse fringe from the ${\it slow}$ phonons corresponding to mode shape D in Fig. 2(d) that are sustained by the anti-crossed 3rd branch in addition to the ${\it fast}$ pulse fringe originating from mode B as seen in the other 2 devices.}
\label{fig 5}
\vspace{-0.7cm}
\end{center}
\end{figure*}

\begin{figure*}[ht]
\begin{center}
\vspace{-2.0cm}\hspace{8cm}
\includegraphics[scale=0.5]{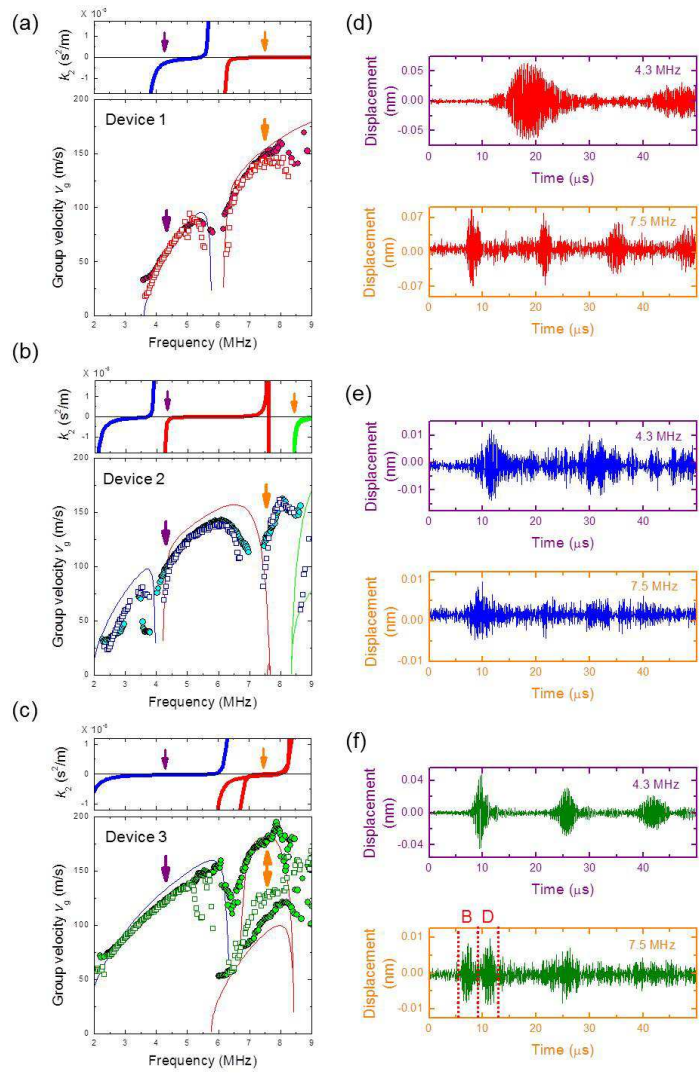}
\vspace{-0.cm}
\caption{(Color online) (a)-(c) The frequency dependence of the group velocity (lower panel) in devices 1, 2 and 3 extracted from the measurements shown in Figs. 5(a)-(c) where the circles and squares correspond to 2 $\mu$s and 8 $\mu$s pulses respectively. The solid lines depict the theoretical group velocities extracted from the FEM derived bandstructure detailed in Figs. 2(a)-(c) with the same color coding enabling the different branches to be identified. For frequencies above 6 MHz in (c), 2 different group velocities can be identified via the shorter pulse data corresponding to the 2 different wavevector phonon modes (B and D in Fig. 2(d)) in the 3rd dispersion branch. The upper panels show the theoretical GVD coefficient extracted from the inverse 2nd derivative of the FEM derived bandstructure detailed in Figs. 2(a)-2(c) where again the same color coding is used enabling the different branches to be identified. This calculation permits the spectral positions at which GVD is maximized or minimized to be located which results in the temporal waveform of the phonon propagation becoming either broadened or unaffected as it travels down the PnC WG. (d)-(f) The output waveforms at 4.3 MHz and 7.5 MHz as indicated by the purple and orange arrows in (a)-(c) respectively at which the GVD is maximized or minimized depending on the GVD coefficient in a given PnC WG.}
\label{fig 6}
\vspace{-0.7cm}
\end{center}
\end{figure*}

\end{document}